\documentclass[pdflatex,sn-nature]{sn-jnl}

\usepackage{longtable}
\usepackage{booktabs}
\usepackage{array}
\usepackage{tabularx}
\usepackage{array}

\usepackage{romannum}
\usepackage{hyperref} 
\usepackage{tcolorbox} 
\tcbuselibrary{breakable}
\usepackage{tabularx}
\usepackage{tcolorbox}
\usepackage{listings}
\usepackage[table]{xcolor}
\usepackage{soul}
\usepackage{booktabs}
\usepackage{multirow}
\usepackage{listings}
\usepackage{placeins} 

\usepackage{float} 
\usepackage{booktabs}
\usepackage{hyperref}
\usepackage{graphicx}   
\usepackage{adjustbox} 
\usepackage{array,tabularx,makecell}
\newcolumntype{Y}{>{\centering\arraybackslash}p{1.7cm}}
\usepackage{tabularx}
\usepackage{graphicx}%
\usepackage{multirow}%
\usepackage{amsmath,amssymb,amsfonts}%
\usepackage{amsthm}%
\usepackage{mathrsfs}%
\usepackage[title]{appendix}%
\usepackage{xcolor}%
\usepackage{textcomp}%
\usepackage{manyfoot}%
\usepackage{booktabs}%
\usepackage{algorithm}%
\usepackage{algorithmicx}%
\usepackage{algpseudocode}%
\usepackage{listings}%
\theoremstyle{thmstyleone}%
\theoremstyle{thmstyletwo}%

\theoremstyle{thmstylethree}%

\begin{document}

\title[MDAgent2]{MDAgent2: Large Language Model for Code Generation and Knowledge Q\&A in Molecular Dynamics}


\author[1,2]{\fnm{Zhuofan} \sur{Shi}}
\equalcont{These authors contributed equally to this work.}

\author[3]{\fnm{Hubao} \sur{A}}
\equalcont{These authors contributed equally to this work.}

\author[4]{\fnm{Yufei} \sur{Shao}}
\equalcont{These authors contributed equally to this work.}


\author[1,2]{\fnm{Dongliang} \sur{Huang}}

\author[1,2]{\fnm{Hongxu} \sur{An}}

\author[1,2]{\fnm{Chunxiao} \sur{Xin}}
\author[1,2]{\fnm{Haiyang} \sur{Shen}}
\author[1]{\fnm{Zhenyu} \sur{Wang}}
\author[6]{\fnm{Yunshan} \sur{Na}}
\author[1,2]{\fnm{Gang} \sur{Huang}}
\author*[1,2]{\fnm{Xiang} \sur{Jing}}\email{jingxiang@pku.edu.cn}

\affil[1]{Peking University}
\affil[2]{National Key Laboratory of Data Space Technology and System}
\affil[3]{The Hong Kong University of Science and Technology}
\affil[4]{Liaoning Technical University}
\affil[6]{Wenjing Future Lab (Beijing) Technology Co., Ltd}


\abstract{
Molecular dynamics (MD) simulations are essential for understanding atomic-scale behaviors in materials science, yet writing LAMMPS scripts remains highly specialized and time-consuming tasks. Although Large Language Models (LLMs) show promise in code generation and domain-specific question answering, their performance in MD scenarios is limited by scarce domain data, the high deployment cost of state-of-the-art LLMs, and low code executability. 
Building upon our prior MDAgent, we present MDAgent2, the first end-to-end framework capable of performing both knowledge Q\&A and code generation within the MD domain. We construct a domain-specific data-construction pipeline that yields three high-quality datasets spanning MD knowledge, question answering, and code generation. Based on these datasets, we adopt a three stage post-training strategy—continued pre-training (CPT), supervised fine-tuning (SFT), and reinforcement learning (RL)—to train two domain-adapted models, MD-Instruct and MD-Code. Furthermore, we introduce MD-GRPO, a closed-loop RL method that leverages simulation outcomes as reward signals and recycles low-reward trajectories for continual refinement. 
We further build MDAgent2-RUNTIME, a deployable multi-agent system that integrates code generation, execution, evaluation, and self-correction. Together with MD-EvalBench proposed in this work, the first benchmark for LAMMPS code generation and question answering, our models and system achieve performance surpassing several strong baselines. 
This work systematically demonstrates the adaptability and generalization capability of large language models in industrial simulation tasks, laying a methodological foundation for automatic code generation in AI for Science and industrial-scale simulations. URL: \url{https://github.com/FredericVAN/PKU_MDAgent2}
}




\maketitle
\section{Introduction}\label{sec1}
Molecular dynamics simulation~\cite{plimptonfast} has become a core tool for exploring material and molecular behavior at the atomic scale \cite{hollingsworth2018molecular}.
With specialized simulation platforms such as LAMMPS, researchers can model a wide range of physical processes—from crystal construction to thermal conductivity analysis \cite{understandingMS}.
However, these simulations typically involve complex modeling scripts, strict physical constraints, and highly structured input formats, which demand strong domain expertise and extensive hands-on experience \cite{MDtheoretical}.
In addition, LAMMPS workflows often require a large amount of repetitive manual work, becoming a bottleneck that limits research efficiency\cite{MDML}.

In recent years, the rapid development of LLMs such as GPT \cite{achiam2023gpt}, Qwen \cite{yang2025qwen3}, and DeepSeek \cite{liu2024deepseek} has driven breakthrough progress in the emerging field of LLMs for Science \cite{Zhang2025Exploringtheroleoflargelanguagemodelsinthescientificmethod}.
Existing studies have shown that large language models (LLMs) are accelerating knowledge discovery and dissemination in materials science \cite{miret2025enabling}. For example,
Jablonka et al.\cite{jablonka202314} showcased multiple LLM applications in materials science and chemistry, while subsequent work\cite{brown2024leveraging_llm_predictive_chemistry} emphasized their potential in predictive chemistry. ChemLLM \cite{ChemLLM} fine-tunes LLMs for chemical materials, and MatterGen \cite{MatterGen} employs generative modeling for inorganic material design. ChemCrow \cite{ChemCrow} leverages API tools to tackle chemistry-related problems, and HoneyComb \cite{HoneyComb} explores LLM-based agent systems for materials science, albeit without fine-tuning. Similarly, ChatMOF \cite{ChatMOF} uses AI systems to predict and generate metal--organic frameworks. MDCrow \cite{campbell2025mdcrow} uses prompt engineering and tool integration for automated MD workflows, and MDAgent~\cite{shi2025fine} introduces a fine-tuned large language model based molecular dynamics agent for code generation to obtain material thermodynamic parameters. Jacobs et al.\cite{jacobs2025developing} improved ORCA input generation via prompt engineering and lightweight synthetic fine-tuning. Dong et al.\cite{dong2025fine} fine-tuned Qwen2.5-7B to generate OpenFOAM CFD configs from natural language using 28k prompt–code pairs. Zhang et al.\cite{zhang2025large} used LLMs to accelerate organic chemistry synthesis, achieving significant advances in active learning–based experimental optimization. Collectively, these studies validate the growing potential of large language models in materials science and the broader AI for Science paradigm.

Although LLMs have demonstrated remarkable potential in scientific computing, adapting LLMs to highly specialized tasks such as LAMMPS-based MD simulations remains challenging. 
In summary, the main challenges include:
\begin{itemize}
\item \textbf{Domain data scarcity and high construction difficulty}:
In materials science research, AI-driven approaches increasingly demand large-scale, high-quality datasets. However, the current materials data ecosystem still suffers from scarcity, fragmentation, inconsistent formats, limited accessibility, and uneven data quality — issues that are particularly pronounced in text-to-code generation tasks. Although several databases and platforms exist, their coverage remains limited and data distribution scattered, making them insufficient to directly support domain-specific post-training of LLMs. At present, there is a lack of high-quality datasets and systematic construction methodologies specifically tailored for molecular dynamics.

\item \textbf{Lack of evaluation benchmarks for LAMMPS}:
Although materials science benchmarks such as DiSCoMaT \cite{gupta2023discomat}, MaScQA \cite{zaki2024mascqa}, ChemBench \cite{walker2010chembench}, and ChemIQ \cite{runcie2025assessing_chem_intelligence} have been developed, these benchmarks provide limited coverage of molecular dynamics simulations and LAMMPS code generation, making it difficult to comprehensively evaluate the capabilities of LLMs in this field.

\item \textbf{Insufficient research on code generation and lack of closed-loop optimization}:
Most existing studies focus on text-to-text generation, with limited exploration of text-to-code generation — particularly for industrial and scientific simulation code in the materials science domain. Many existing methods remain at the one-shot generation stage, lacking closed-loop mechanisms for automated execution, evaluation, and self-correction, which hinders continuous improvement in the quality of generated code\cite{shi2025fine}.

\item \textbf{High cost and limited deployability of large-scale SOTA LLMs}: 
Although state-of-the-art general-purpose LLMs exhibit certain capabilities in MD-related tasks, their practical use remains constrained. Proprietary commercial models (e.g., GPT-5.1) cannot be deployed locally due to closed-source restrictions, while ultra-large open-source models (e.g., Qwen3-235B) impose prohibitive inference costs. These limitations highlight the necessity of developing lightweight, domain-specialized LLMs that can be deployed efficiently on local systems while maintaining strong code generation and reasoning performance.

\end{itemize}

\begin{figure}[H]
\centering
\includegraphics[width=1.0\linewidth]{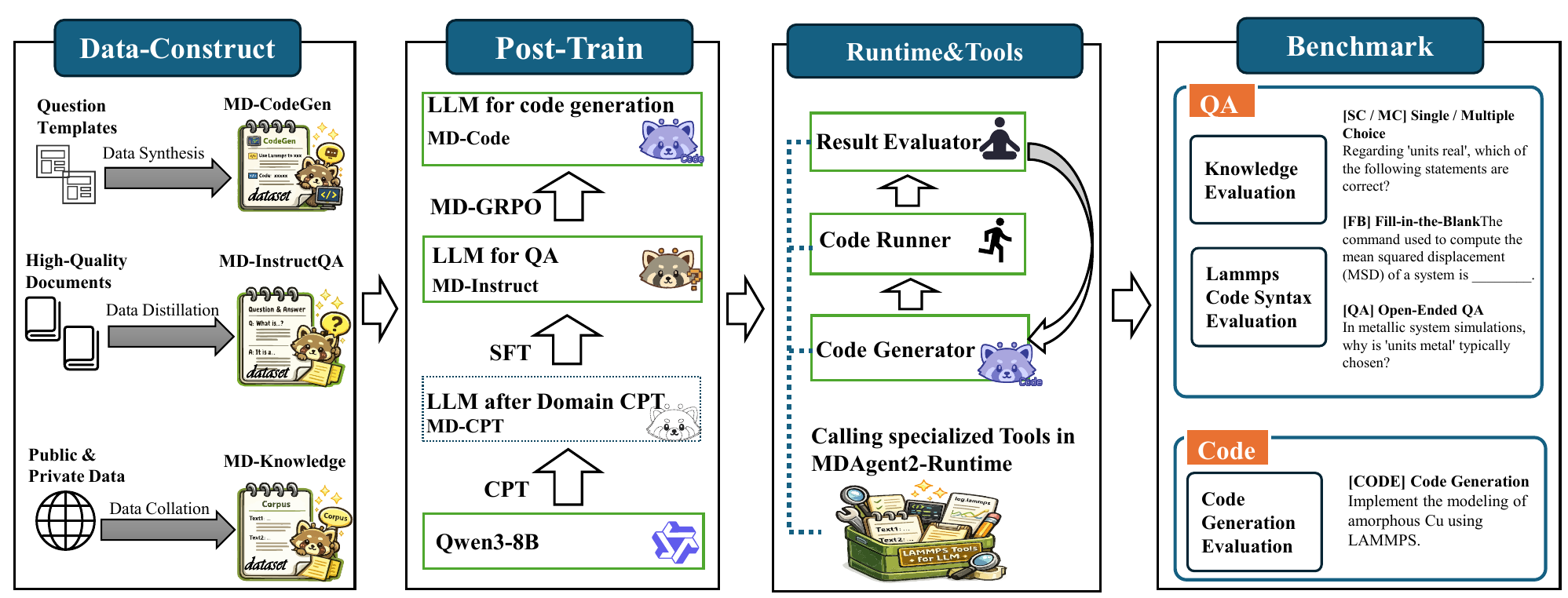}
\caption{Overall workflow of the proposed MDAgent2 framework, integrating data construction, model training, multi-agent runtime, and evaluation.}
\label{fig:mdagent2-workflow}
\end{figure}

To address these challenges, this paper proposes \textbf{MDAgent2}, the first framework designed for code generation and knowledge Q\&A in Molecular Dynamics, as shown in Fig.~\ref{fig:mdagent2-workflow}.

At the data level, we design a structured data construction pipeline that produces three high-quality datasets: \emph{MD-Knowledge} (for domain adaptive pretraining), \emph{MD-InstructQA} (for instruction fine-tuning), and \emph{MD-CodeGen} (for simulation code generation). These datasets cover diverse material systems, physical conditions, and simulation tasks, filling a long-standing gap in high-quality data resources for the molecular dynamics domain.

At the methodological training level, prior studies have shown that continued pre-training (CPT) and supervised fine-tuning (SFT) are key techniques for domain adaptation and task alignment of LLMs\cite{xie2024efficient,lu2025fine}. Meanwhile, a growing body of research on code LLMs indicates that incorporating end-to-end reinforcement learning (RL) with execution-based feedback can further improve code generation quality and reliability\cite{wang2024enhancing,gehring2024rlef}. In contrast, prior MDAgent\cite{shi2025fine} mainly relied on SFT and did not incorporate execution-feedback-driven RL into the training loop, leaving room for further improvement.

Motivated by these findings, we conduct a three-stage post-training pipeline: CPT, SFT, and RL. Specifically, CPT injects molecular dynamics and LAMMPS-related knowledge from large-scale unlabeled corpora, enhancing domain representations as well as the model's mastery of specialized terminology and structured formats. We then perform SFT on high-quality instruction data to align the model with MD task requirements.

Moreover, we introduce \emph{MD-GRPO}, a closed-loop RL framework built upon GRPO\cite{shao2024deepseekmath}. After generating simulation code, the system automatically executes it, evaluates the outcomes, and uses the resulting scores as reward signals to optimize the policy. In addition, we propose a low-reward trajectory recycling mechanism to continually refine generation strategies, effectively improving the executability and physical correctness of the generated code.

At the model level, we train two specialized models based on the Qwen3 series: \emph{MD-Instruct} (domain understanding model) and \emph{MD-Code} (simulation code generation model), which substantially enhance the model's capability to generate accurate and executable codes within materials science contexts.

At the system level, we implement a deployable multi-agent runtime system, MDAgent2-RUNTIME, which enables a fully automated workflow from natural-language task descriptions to industrial-grade simulation code generation, execution, evaluation, and self-correction.

At the evaluation level, we establish the first benchmark for molecular dynamics question answering and code generation, MD-EvalBench. Experimental results demonstrate that our LLM exhibits strong question-answering and code generation capabilities compared to selected baselines. Furthermore, MDAgent2-RUNTIME effectively enhances the model’s ability to generate correct and executable LAMMPS code.


\section{Results}\label{sec2}
\subsection{Evaluation Benchmark for Molecular Dynamics: \emph{MD-EvalBench}}

To comprehensively evaluate the performance of LLMs in the MD domain, we construct an integrated benchmark suite named MD-EvalBench, consisting of three complementary datasets: MD-KnowledgeEval, LAMMPS-SyntaxEval, and LAMMPS-CodeGenEval.
These datasets jointly assess model capabilities from three key perspectives — theoretical understanding, syntactic comprehension, and code generation.

We collaborated with domain experts to design MD-KnowledgeEval and LAMMPS-SyntaxEval.
Both datasets share a unified question structure encompassing four types: single-choice, multiple-choice, fill-in-the-blank, and short-answer questions.

MD-KnowledgeEval: Theoretical Knowledge Assessment in Molecular Dynamics. 
This dataset evaluates the model's understanding of fundamental concepts, simulation principles, and thermodynamic systems in molecular dynamics.
It contains a total of 336 expert-curated questions covering topics such as interatomic potentials, integration algorithms, equilibrium conditions, and statistical ensembles.

LAMMPS-SyntaxEval: Command and Syntax Understanding Assessment. 
This dataset focuses on the model's comprehension of LAMMPS scripting — including command usage, syntax rules, parameter structures, and functional modules — thereby measuring its ability to interpret and reason about simulation scripts.
It comprises 333 questions designed to test practical command-level proficiency.

LAMMPS-CodeGenEval: Automatic Code Generation Quality Assessment. 
This dataset assesses the model’s ability to automatically generate executable LAMMPS scripts from natural language task descriptions.
Each sample describes a user-defined simulation objective, and the model is required to produce corresponding runnable code.
Generated scripts are then evaluated through structural and functional scoring criteria to quantify generation accuracy and execution reliability.

\subsection{Experimental Setup}
\paragraph{Evaluation Protocol}
To ensure a fair comparison, all experiments are repeated three times and 
the average results are reported. 
We adopt two primary metrics. 
\emph{Exec-Success@$k$} measures the proportion of tasks for which at least one of the
$k$ generated candidates can be successfully executed in LAMMPS.
\emph{Code Human Score} is a subjective rating in the range $[0,10]$ assigned by
domain experts based on readability, robustness, and physical correctness.

\paragraph{Baselines and Compared Models}
We compare MDAgent2 against a range of representative systems and models:
MDAgent~\cite{shi2025fine}, a multi-agent framework using a 
generate–evaluate–rewrite loop; 
Direct Prompting, which generates LAMMPS scripts via a single prompt without tool integration, agent orchestration, or execution feedback. Qwen3~\cite{yang2025qwen3}, a general-purpose LLM family released by Alibaba.

\subsection{Evaluation of QA Ability}
\begin{table}[ht]
\centering
\small
\caption{Performance comparison across three benchmarks. Total Score denotes the aggregated QA performance over all question types. $\Delta$ indicates the absolute improvement over the 8B baseline (Qwen3-8B). Boldface highlights the best result in each column.}
\label{tab:exp1}

\begin{tabular}{l c cc cc cc}
\toprule
\textbf{MODEL} 
& \textbf{Size / Access}
& \multicolumn{2}{c}{\textbf{Overall Avg}}
& \multicolumn{2}{c}{\textbf{Knowledge}}
& \multicolumn{2}{c}{\textbf{Syntax}} \\
\cmidrule(lr){3-4}\cmidrule(lr){5-6}\cmidrule(lr){7-8}
& 
& \textbf{All} & \textbf{$\Delta$}
& \textbf{All} & \textbf{$\Delta$}
& \textbf{All} & \textbf{$\Delta$} \\
\midrule

Qwen3-max
& Large / Closed
& \textbf{82.49} & +11.99
& \textbf{86.57} & +11.42
& \textbf{78.40} & +12.56 \\

Qwen3-32b
& 32B / Open
& 77.34 & +6.84
& 81.94 & +6.79
& 72.74 & +6.90 \\

\rowcolor[RGB]{225,235,245}
MD-Instruct-8B
& 8B / Open
& 74.67 & +4.17
& 76.89 & +1.74
& 72.45 & +6.61 \\

Qwen-flash
& Large / Closed
& 73.47 & +2.97
& 78.64 & +3.49
& 68.30 & +2.46 \\

Qwen3-14b
& 14B / Open
& 72.91 & +2.41
& 77.90 & +2.75
& 67.92 & +2.08 \\

\rowcolor[gray]{0.92}
Qwen3-8b
& 8B / Open
& 70.50 & 0.00
& 75.15 & 0.00
& 65.84 & 0.00 \\

\bottomrule
\end{tabular}
\end{table}

We evaluate the QA capability using MD-KnowledgeEval and LAMMPS-SyntaxEval.
The QA results in Table~\ref{tab:exp1} reveal two clear findings.

\textbf{MD-Instruct-8B shows competitive performance despite its smaller size.}
With domain-specific post-training, MD-Instruct-8B attains an average score of 74.67, 
surpassing Qwen-Flash and Qwen3-14B, and substantially narrowing the gap to the 
much larger Qwen3-32B. This suggests that the model has acquired a 
grasp of MD-related domain knowledge, providing a solid foundation for subsequent 
code-generation tasks. We attribute this advantage to domain-specific post-training: CPT injects MD knowledge and strengthens the model's understanding of MD concepts, while SFT further consolidates this knowledge through supervised MD question answering and promotes faithful, well-structured responses. 

\textbf{Qwen3-Max provides the strongest overall performance.}
Qwen3-Max achieves the highest average score of 82.49, demonstrating that large-scale 
SOTA LLMs still retain strong generalization ability in both MD theoretical 
knowledge and LAMMPS syntax understanding.

\subsection{Evaluation of Code Generation Ability}

\begin{figure}[ht]
    \centering
    \includegraphics[width=1.0\linewidth]{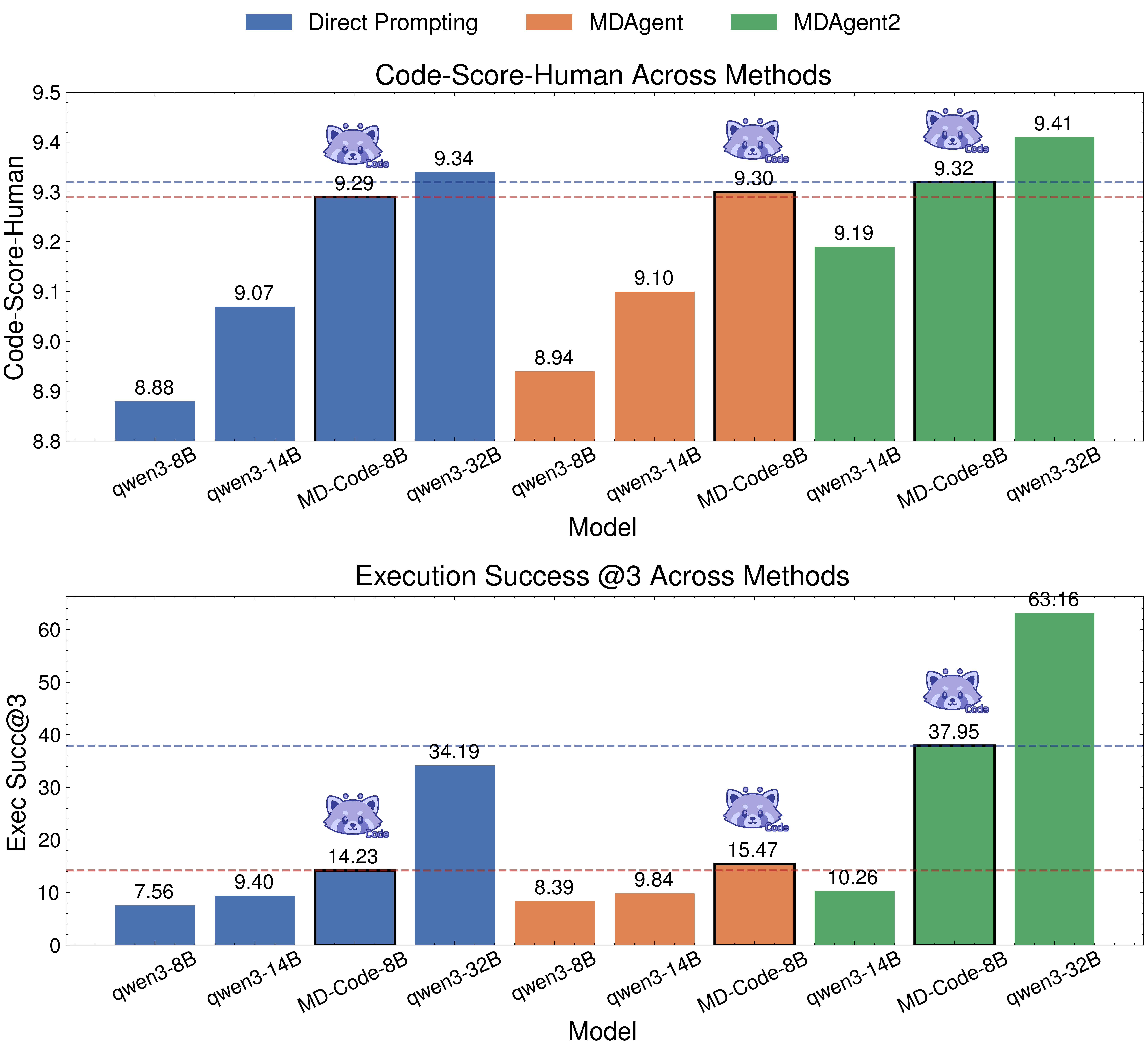}
    \caption{Comparison of Code-Score-Human and Execution Success@3 across all methods.}
    \label{fig:Exp2}
\end{figure}
We evaluate the code generation capability using the LAMMPS-CodeGenEval benchmark.  
The results in Fig.~\ref{fig:Exp2} highlight several important observations.

\textbf{MDAgent2-RUNTIME significantly improves code-generation performance.}
Taking MD-Code-8B as an example, enabling the RUNTIME loop boosts \textit{ExecSucc@3} from 14.23\,\% to 37.95\,\%, and slightly improves \textit{Code-Score-Human} from 9.29 to 9.32.
These results highlight the value of iterative evaluation and self-correction in a multi-agent system for improving the reliability and executability of generated LAMMPS scripts.
Notably, the gains also rely on the task-specific tools we design for LAMMPS, which provide actionable feedback signals during the runtime loop.

\textbf{MDAgent2-RUNTIME outperforms the MDAgent~\cite{shi2025fine} multi-agent framework.}
Across the three backbone models evaluated, MDAgent2 consistently achieves better results than the MDAgent framework.
We attribute the improvement to two key differences: (i) MDAgent2 incorporates dedicated, LAMMPS-tailored tools; and (ii) it leverages feedback from actual LAMMPS execution, which was not available in MDAgent.

\textbf{Domain-specific post-training of MD-Code-8B is crucial.}
After post-training, MD-Code-8B attains a \textit{Code-Score-Human} of 9.29 even under the \textit{Direct Prompting} generation setting, indicating a substantial improvement over the Qwen3-8B baseline.
We attribute this gain to the full post-training stack: the first two stages strengthen domain knowledge and improve the model’s understanding of LAMMPS syntax and knowledge, while the MD-GRPO RL stage further enhances end-to-end code generation capability by optimizing directly against execution-based feedback. Consequently, domain-specific post-training substantially improves the model’s ability to produce accurate and executable LAMMPS scripts, providing a stronger base model for agentic runtime optimization.

\section{Methods}\label{sec11}

\subsection{Construction of Training Datasets}
To support multi-task capability training and high-quality evaluation, three datasets were constructed in this study, respectively used for incremental pre-training, fine-tuning, and reinforcement learning.

\paragraph{\emph{MD-Knowledge}: A Pretraining Corpus for Molecular Dynamics} 
We organized a team of domain experts to systematically collect thousands of high-quality molecular-dynamics–related papers, textbooks, technical documents, and public manuals as the primary sources for corpus construction.
Following best practices established in large-scale pretraining corpora such as Lee et al.\cite{lee2022deduplicating} and Soldaini et al.\cite{soldaini-etal-2024-dolma}, we designed and implemented a multi-stage data-cleaning pipeline to ensure that the resulting corpus is high-quality, low-redundancy, and minimally sensitive for large language model training:

\begin{enumerate}
    \item \textbf{Useless Text Filtering} 
    The raw text is first pre-screened to remove empty strings, corrupted content, and samples with excessively short lengths, thereby improving the baseline quality of the corpus.

    \item \textbf{Regex-based Content Removal} 
    Regular expressions are used to automatically identify and remove potential sensitive information such as DOI numbers, URLs, and email addresses.

    \item \textbf{Approximate Deduplication via MinHash and LSH} 
    The MinHash technique is used to represent textual data as approximate sets, and combined with Locality Sensitive Hashing (LSH) to perform large-scale efficient text deduplication.  
    This effectively reduces semantically similar but textually varied redundant data.

    \item \textbf{Semantic Deduplication via Embedding Similarity} 
    Each text sample is encoded using the pretrained sentence embedding model to obtain its semantic representation.  
    Cosine similarity between samples is then computed to further identify and remove semantically duplicate or highly similar content, achieving fine-grained deduplication.

    \item \textbf{High-Quality Sample Filtering via LLM Evaluation} 
    The \texttt{deepseek-chat} model is employed as an automatic quality evaluator.  
    By designing appropriate prompts to guide the model’s assessment, the quality of each text is evaluated across multiple dimensions—including linguistic clarity, logical coherence, and information density—to select high-quality samples suitable for model training.
\end{enumerate}

\paragraph{\emph{MD-InstructQA}: A Domain-Specific Question–Answer Dataset for Molecular Dynamics}

Based on the original corpus resources collected above, we further constructed a domain knowledge question answering dataset \texttt{MD-InstructQA} for the instruction fine-tuning stage through a set of automated semantic extraction and question-answer generation processes. The dataset contains two parts: one with reasoning and one without reasoning.

First, we convert original PDF documents into Markdown through a multi-stage workflow encompassing filtering, metadata extraction, layout analysis, structured content extraction, specialized recognition of elements such as formulas and tables, reading order sorting, and new format generation. This enables seamless utilization of the content in downstream tasks like LLM fine-tuning. We then employed a structure-sensitive hybrid chunking process to flexibly segment the converted Markdown documents into semantically coherent text chunks. This process, which takes titles, paragraphs, and catalogs into account, balances automation with manual control, ensuring the resulting chunks are compatible with LLM context windows while preserving semantic integrity. Next, we generate a semantic domain label tree based on the semantic chunks, and use deliberately tailored prompts to generate questions that accurately reflect the key information in the label tree—thereby ensuring diversity in question styles. Furthermore, to generate answers that are faithful to the source text—preserving key factual details and semantic meaning—we employed knowledge-enhanced prompting with a selectable chain-of-thought process. This approach prevents hallucination and ensures answers remain semantically aligned with the source content. Lastly, we paired the generated questions and answers into QA pairs and formatted them into the Alpaca schema to streamline the LLM fine-tuning workflow. 

\paragraph{\emph{MD-CodeGen}: A Code Generation Dataset for Molecular Dynamics Simulations}

This dataset consists of paired data samples in the form of (task description, LAMMPS code).
On one hand, domain experts manually collected and constructed high-quality examples.
On the other hand, we designed an automated task modeling approach to generate large-scale data by combining multiple key elements such as “material system,” “research objective,” and “simulation conditions.”
Natural-language task descriptions are created through structured templates and then manually reviewed for correctness before being used as inputs.
Subsequently, based on the proposed MDAGENT2-RUNTIME with SOTA LLMs produce physically meaningful LAMMPS scripts.
Finally, simulation experts with extensive materials science experience carefully review and refine each generated sample, yielding a high-quality synthetic dataset.

\subsection{Post-Training for MD-LLM}

To enhance the model’s understanding and reasoning capabilities in molecular dynamics, we conduct a three-stage post-training process on the Qwen3-8B backbone using our constructed datasets: MD-Knowledge, MD-InstructQA, and MD-CodeGen. The resulting model, denoted as MD-Instruct-8B and MD-Code-8B.

\subsubsection{CPT and SFT}

\textbf{Continual Pretraining (CPT).}
In this stage, we perform incremental domain-adaptive pretraining by mixing the MD-Knowledge corpus with general-domain data. This process enhances the model’s representation of materials terminology, simulation workflows, and structural concepts, while preserving its general linguistic competence, thereby providing a solid semantic foundation for subsequent instruction tuning.

\textbf{Supervised Fine-Tuning (SFT).}
Next, we construct an instruction-aligned training set by mixing MD-InstructQA with general instruction data. The model is fine-tuned under a supervised objective to align its responses with domain reasoning patterns and improve factual accuracy. Additionally, a curated subset of MD-CodeGen samples is introduced as a cold-start seed to expose the model to task–code mappings at an early stage.

After these two stages, we obtain a domain-aware language model, \texttt{MD-Instruct-8B}, capable of comprehending and answering molecular dynamics–related questions with strong domain fidelity, laying the groundwork for reinforcement learning in the subsequent MD-GRPO stage.

\subsubsection{MD-GRPO}
GRPO~\cite{shao2024deepseekmath}, a widely adopted algorithm for reinforcement learning in LLMs~\cite{zhang2025surveyreinforcementlearninglarge}, has already demonstrated substantial value, with domain-specific variants such as Med-R1~\cite{lai2025med} and QoQ-Med~\cite{dai2025qoq} achieving strong performance in vertical applications.
Building upon GRPO, we develop a reinforcement learning framework named \texttt{MD-GRPO} to train LLMs for generating executable and goal-achieving lammps scripts from natural-language task descriptions.
Given an input query, the policy model produces multiple candidate codes, which are executed in a molecular dynamics simulation environment and evaluated by a reward model based on code and results.Then policy optimization, forming an execution-driven closed-loop learning process.

\begin{figure}
    \centering
    \includegraphics[width=1.0\linewidth]{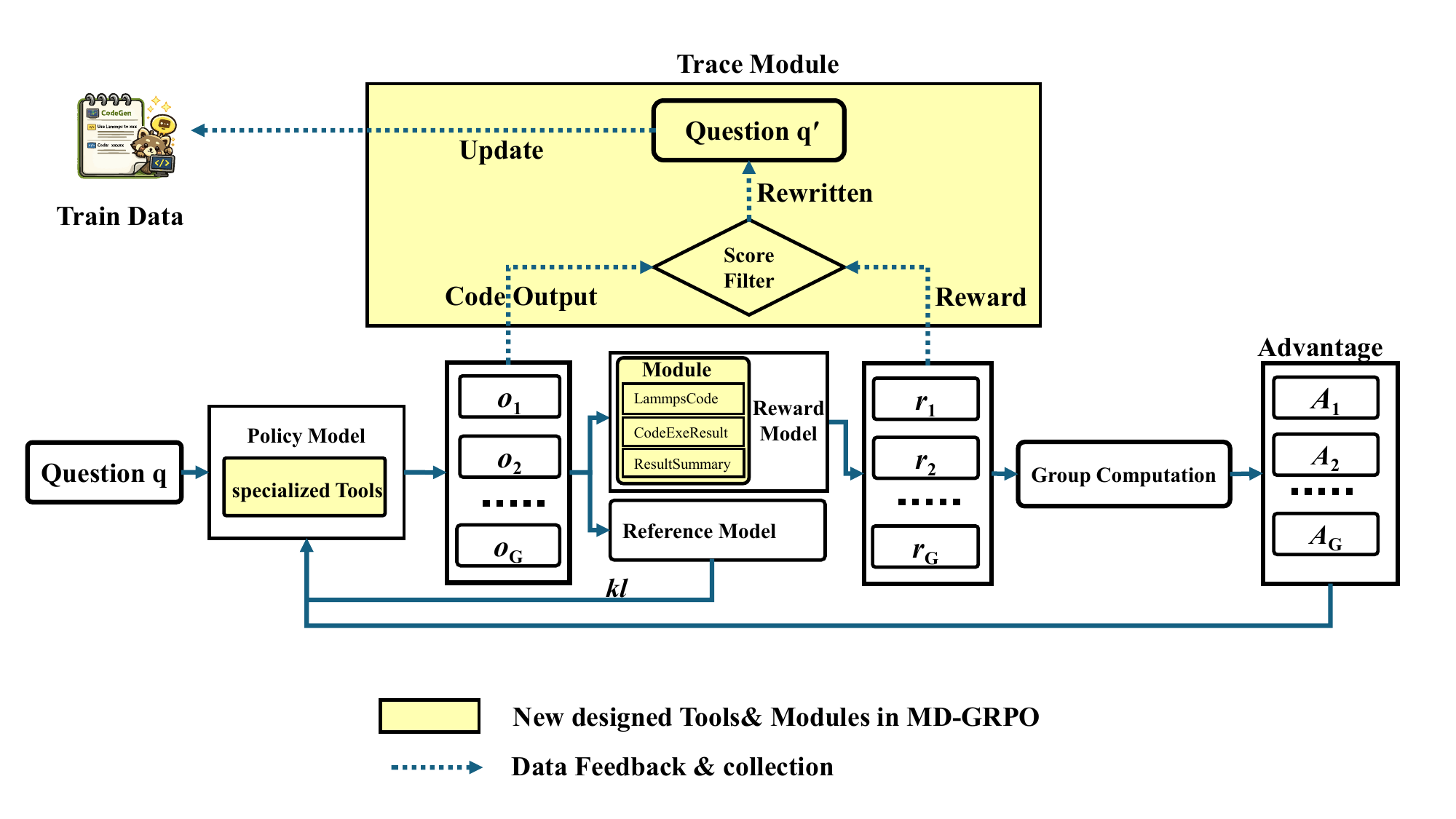}
    \caption{Overview of the MD-GRPO. The framework integrates code generation, simulation execution, and trajectory rewriting to enable execution-driven optimization for molecular dynamics tasks.}
    \label{fig:MD-Instruct-GRPO}
\end{figure}

To further enhance the model’s self-correction capability, MD-GRPO introduces a failure-aware trace-and-rewrite mechanism(Trace Module in Fig.~\ref{fig:MD-Instruct-GRPO}) into each training iteration.
When a generated code fails to execute or receives a low evaluation score, the system records the entire trajectory, including the low-scoring code and its corresponding evaluation feedback, and rewrites the original \textit{\textbf{q}} into a refined form \textit{\textbf{q'}}.
The rewritten query is then added to the training data pool, enabling the model to iteratively correct previously observed failure patterns.

Through this design, the framework extends conventional single-step reinforcement learning to trajectory-aware optimization that more closely reflects real runtime behavior, where multiple rounds of code refinement are often necessary.
As a result, the model not only learns \emph{how to generate valid simulation code}, but also progressively understands \emph{why earlier attempts failed and how to correct them}, forming an automated, reward-driven self-improvement loop.

\paragraph{Total Reward}

\begin{equation}
\label{eq:reward-total-extended}
\mathcal{R}_{\text{total}} = \lambda_1 \mathcal{R}_{\text{format}} + \lambda_2 \mathcal{R}_{\mathrm{correct}},
\end{equation}

where:
\begin{itemize}
  \item \(\mathcal{R}_{\text{format}} \in \{0, 1\}\): \textbf{Format reward}, which equals 1 only when the generated output satisfies the required formatting constraints; otherwise, it is 0.  
  \item \(\hat{R}_{\mathrm{correct}} \in [0,1]\): \textbf{Correctness reward}, representing the normalized multidimensional score that evaluates the quality of the generated LAMMPS script.  
  \item  \(\lambda_1\) =1 and \(\lambda_2\)  =5 by default.
\end{itemize}

\paragraph{Format Reward}
To ensure the model adheres to a structured reasoning-and-answering protocol, we design a binary format reward $\mathcal{R}_{\text{format}} \in \{0, 1\}$ that imposes both syntactic and semantic constraints on the output. Specifically, the model is expected to first produce a reasoning trace enclosed within \texttt{<think>} tags, followed by a final answer enclosed in \texttt{<answer>} tags.

Moreover, the content inside the \texttt{<answer>} block must be a valid JSON object that can be parsed without error, and must contain \emph{exactly} the required set of fields as specified by the task—no missing fields, and no extraneous ones.

\[
\mathcal{R}_{\text{format}} =
\begin{cases}
1, & \parbox[t]{0.75\linewidth}{\raggedright if \texttt{<think>} and \texttt{<answer>} appear in order, and \texttt{<answer>} contains a valid and structurally correct JSON} \\
0, & \text{otherwise}
\end{cases}
\]

This reward encourages the model to output not only interpretable reasoning steps, but also task-specific structured results that can be programmatically consumed and verified. During training, we implement the format check via regular expressions combined with strict JSON schema validation.

\paragraph{Correct Reward}
Based on the evaluation dimensions proposed in MDAgent~\cite{shi2025fine}, we invited domain experts to further refine and expand the framework into eight dimensions that comprehensively capture the essential aspects of lammps script quality assessment(see Appendices~\ref{sec:eval details} for details). 
Building on these dimensions, we design a hybrid reward that integrates both additive bonus signals and penalty signals, enabling a balanced evaluation of LAMMPS script quality across all key factors.
\begin{equation}
\label{eq:reward}
R_{\text{raw}}
= \sum_{k=1}^{K} w_k^{+} B_k
- \sum_{m=1}^{M} w_m^{-} P_m,
\end{equation}
and we map it to $[0,1]$ via a clip-and-rescale operation:
\begin{equation}
R_{\mathrm{correct}}
=\mathrm{scale}\!\left(\mathrm{clip}(R_{\text{raw}},R_{\min},R_{\max})\right).
\end{equation}

where
\begin{itemize}
    \item $B_k\in\{0,1\}$ is a \textit{bonus indicator} that equals 1 if the $k$-th key module is completed;
    \item $P_m\in\{0,1\}$ is a \textit{penalty indicator} that equals 1 if the $m$-th error type is detected.
    \item $w_k^{+}$ and $w_m^{-}$ are expert-assigned weights reflecting the relative importance of each component.
    \item $R_{\min}$ and $R_{\max}$ denote the minimum and maximum possible values of $R_{\text{raw}}$.
\end{itemize}


\subsection{Runtime Multi-Agent Framework}

\begin{figure}[t]
\centering
\includegraphics[width=1.0\linewidth]{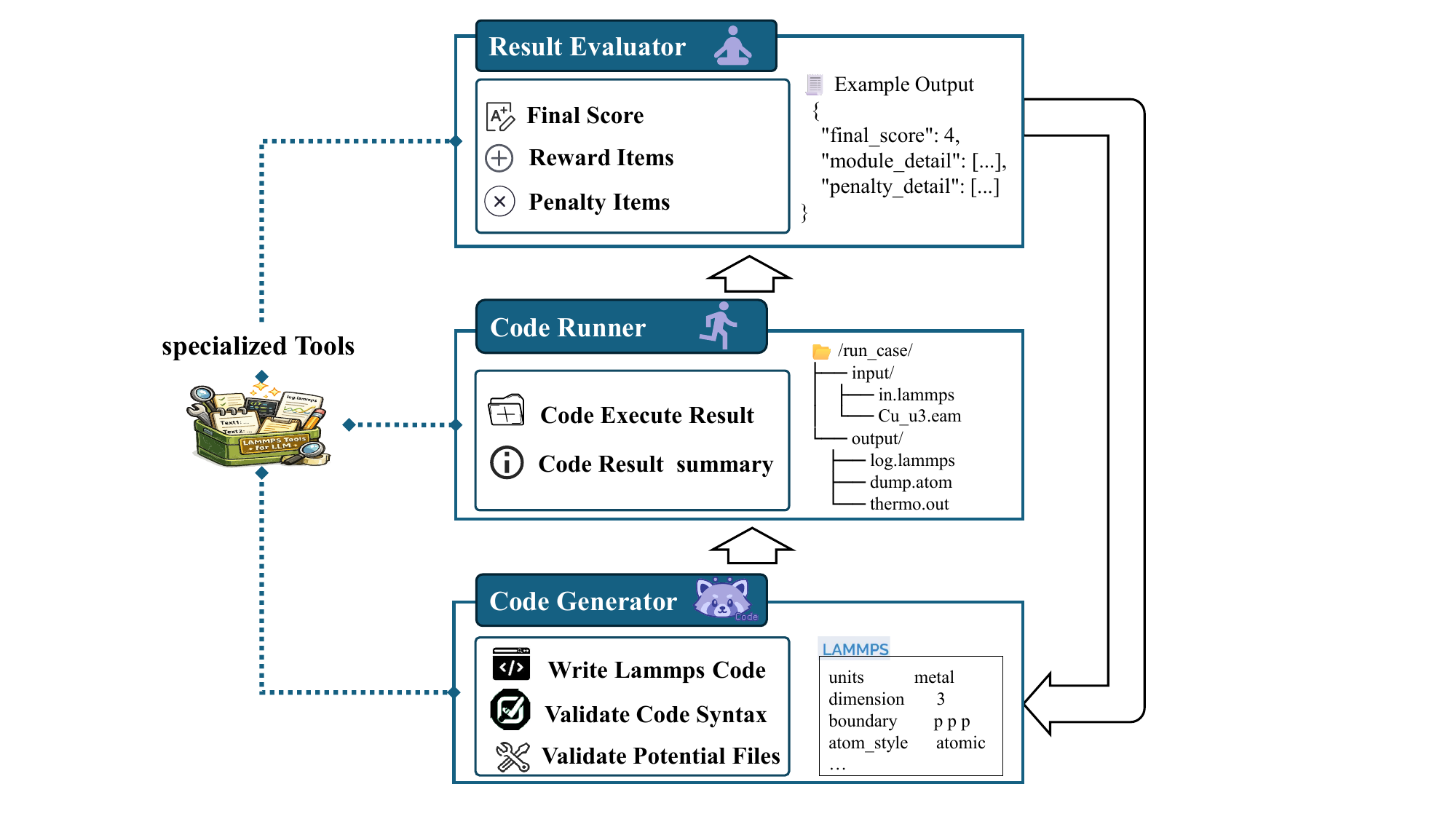}
\caption{Overall architecture of the MDAgent2-RUNTIME multi-agent system.}
\label{fig:mdagent2-runtime}
\end{figure}
After the model training stage, we construct the runtime multi-agent system \texttt{MDAgent2-RUNTIME}, built upon the previous version, MDAgent~\cite{shi2025fine}, as illustrated in Fig.~\ref{fig:mdagent2-runtime}. Once a user provides a natural language task description, the system automatically completes code generation, simulation execution, and result evaluation without requiring any manual intervention.

Additionally, to empower the agents we have developed a suite of MD tools for agent use, as detailed in the Appendix~\ref{sec:tools_details}. A visualization tool is provided to generate thermodynamic curves such as temperature, energy, and pressure versus time, as well as to convert atomic trajectory dump files into GIF animations for intuitive observation. An automated quality evaluation tool is implemented to identify the simulation type and to conduct multi-dimensional assessments based on energy stability, temperature control, pressure consistency, and numerical robustness. We further implement a LAMMPS syntax verification tool that rapidly determines whether a script is executable by performing an actual launch with a timeout mechanism, returning results immediately without waiting for the full simulation to complete.

In addition, based on our experimental findings, LLMs show limited capability in selecting appropriate potential functions. Therefore, we provide a set of potential-related tools. In addition to listing all available local potential files and retrieving information about specific potential files, we also implement a potential file management tool that automatically scans the generated LAMMPS scripts to detect missing local potential files and supplements them as needed to ensure smooth execution. Furthermore, when the potential function provided by the model is incorrect, we provide an automatic recommendation tool that suggests the Top-K most similar potential functions.

The entire system is organized into three functional nodes: \textit{Code Generator}, \textit{Code Runner}, and \textit{Result Evaluator}. 

In Code Generator, the Writer LLM drafts the initial LAMMPS code, which then undergoes two levels of verification by tools calling.
First, the Syntax tools checks for syntax correctness, identifies potential issues, and provides feedback for revision.
Next, the potential-file tools can be invoked,
if the LLM has specified a valid LAMMPS potential file, the tool ensures that the corresponding parameter file is available locally; if the specified potential file does not actually exist, the tool automatically recommends the Top-K most similar LAMMPS potential files.
Based on the feedback from these two checks, the Writer LLM internally revises and regenerates its code iteratively, continuing this loop until either convergence or a predefined iteration limit is reached.

The Code Runner executes the generated LAMMPS code safely within a sandboxed Docker container by calling tools, ensuring environment isolation and reproducibility. Execution results are stored in designated directories, accompanied by an automatically generated summary of results.

The Result Evaluator then analyzes both the input code and the simulation outputs based on the multi-dimensional evaluation criteria proposed in this work, generating a structured score with detailed reward and penalty components. These feedback signals are returned to the Code Generator to guide further refinement; if the final score falls below a predefined threshold, the system automatically initiates a new iteration of code regeneration.

Meanwhile, MDAgent2-RUNTIME supports human-in-the-loop interaction. Users may pause the automated workflow at any stage and provide natural-language feedback or directives. For example, before execution, users can modify the generated code, adjust simulation parameters, or inject domain-specific instructions. This flexible integration of automation and human expertise enables a closed-loop optimization process that continuously improves code quality and simulation reliability.

\section{Discussion}\label{sec12}
In this work, we presented MDAgent2, an end-to-end large language model framework for molecular dynamics knowledge question answering and simulation code generation. 
By systematically addressing domain data scarcity, the lack of executable feedback, and insufficient evaluation standards in existing approaches, MDAgent2 advances MD code generation from a one-shot text-to-code paradigm toward a closed-loop workflow.

Specifically, we constructed a high-quality domain-specific data pipeline spanning MD knowledge, instruction reasoning, and LAMMPS code generation, and trained two lightweight yet effective domain-adapted models, MD-Instruct and MD-Code. 
Building upon these models, we introduced MD-GRPO, a reinforcement learning framework that directly leverages simulation execution outcomes as reward signals, enabling continuous refinement toward executable and physically meaningful MD scripts. 
Furthermore, we implemented MDAgent2-RUNTIME, a deployable multi-agent system that integrates generation, execution, evaluation, and self-correction into an automated simulation pipeline.

Extensive experiments on the proposed MD-EvalBench benchmark demonstrate that MDAgent2 and the MD-series LLMs can effectively address molecular dynamics question answering and simulation code generation tasks.

Finally, while this work focuses on molecular dynamics as a representative domain, the proposed data construction paradigm, execution-aware reinforcement learning, and agentic closed-loop workflow are broadly applicable to other AI-for-Science scenarios involving simulation-driven reasoning and code generation.

\subsection*{Limitation and Outlook}
Currently, the dataset covers simulation tasks including the computation of thermodynamic properties of materials, fluid dynamics simulations, and mechanical property simulations of materials; however, we plan to further expand the range of supported task types in future work.

Future work will explore the integration of multimodal LLMs into the MDAgent framework. The outputs of LAMMPS simulations—such as visualized thermodynamic curves, atomic trajectories, and structural evolution in .gif or .png formats—provide rich visual information that can be incorporated as additional evaluation inputs to improve interpretability and assessment granularity.

Furthermore, the proposed framework demonstrates strong generality and scalability. Therefore, we aim to extend its application to other scientific fields in future work. Its modular design facilitates seamless adaptation not only within molecular dynamics but also across other industrial and materials science simulation domains, paving the way toward a universal AI-driven framework for scientific code generation and autonomous experimentation.

\backmatter

\bmhead{Acknowledgements}
We acknowledge the support of the National Supercomputing Center in Tianjin for providing high-performance computing resources.

\section*{Declarations}

\subsection*{Data and Code Availability}
The data and code will be made publicly available upon acceptance at 
\url{https://github.com/FredericVAN/PKU_MDAgent2}

\bibliography{sn-bibliography}

\begin{appendices}

\section{Tools Details}\label{sec:tools_details}
Table~\ref{tab:mdagent2_tools} provides a comprehensive list of the tools integrated into the MDAgent2 system. Each entry details the tool's name and its specific functionality within the molecular dynamics workflow. Supplementary to the system architecture discussed in Section 3, this table details the specific tools available to the agent.
\begin{table}[h!]
\centering
\renewcommand{\arraystretch}{1.25}
\resizebox{\textwidth}{!}{ 
\begin{tabular}{p{0.3\textwidth} p{0.7\textwidth}}
\toprule
\textbf{Tool Name} & \textbf{Description} \\
\midrule
check\_syntax\_tool & Verifies whether a LAMMPS input script is syntactically correct and executable, using either LLM-based semantic assessment or a dry-run execution test. \\
check\_lammps\_potentials\_tool & Scans referenced interatomic potential files in the input script, checks their local availability, and attempts to download missing files from official LAMMPS sources (failure to download does not imply non-existence). \\
get\_potential\_file\_info\_tool & Retrieves detailed metadata and content information for a specified potential file. \\
list\_available\_potentials\_tool & Lists all potential files available in the local potentials directory. \\
find\_similar\_potentials\_tool & Given a possibly misspelled potential name, searches local potential files and returns the most relevant matches. \\
check\_potential\_real\_exists\_tool & Uses an LLM-based search module or web search API to determine whether a potential file truly exists. Returns \texttt{False} only when confirmed nonexistent. \\
visualize\_tool & Visualizes LAMMPS outputs (e.g., log.lammps, dump.lammpstrj) including trajectory plots and diagnostic curves. \\
evaluate\_log\_quality\_by\_rule\_tool & Performs rule-based and multi-dimensional evaluation of LAMMPS log quality, such as convergence, numerical stability, and error patterns. \\
lammps\_run\_tool & Executes LAMMPS simulations and stores all generated output files in a designated directory. \\
\bottomrule
\end{tabular}
}
\caption{Summary of tool functionalities in the MDAgent2 system.}
\label{tab:mdagent2_tools}
\end{table}
\FloatBarrier

\section{Evaluation Details}\label{sec:eval details}

To ensure a rigorous and standardized assessment of the generated LAMMPS scripts, we established a multi-dimensional correctness evaluation framework, as detailed in Table~\ref{tab:reward-dimensions}. This framework encompasses eight distinct dimensions, ranging from basic Syntax Correctness to high-level Physical Soundness.

Crucially, this rubric serves a dual purpose in our system. First, it provides a unified guideline for human experts to conduct ground-truth evaluations. Second, these exact dimensions and deduction rules are explicitly incorporated into the prompt design for the LLM. By aligning the LLM's scoring instructions with this expert-verified rubric, we ensure that the automated MDAgent Correctness Reward remains consistent with human scientific judgment.
\begin{table*}[htbp]
    \small
  \centering
  \caption{Evaluation dimensions and typical deduction examples for the MDAgent Correctness Reward.}
  \label{tab:reward-dimensions}
  \renewcommand\arraystretch{1.15}
  \resizebox{0.9\textwidth}{!}{ 
  \begin{tabularx}{\textwidth}{c l X X}
    \toprule
    \textbf{ID} & \textbf{Evaluation Dimension} & \textbf{Description} & \textbf{Typical Deduction Example} \\
    \midrule
    1 & \textbf{Syntax Correctness} &
        Whether the script follows valid \texttt{LAMMPS} syntax rules and avoids misspellings or invalid commands. &
        Invoking a non-existent command or having a misspelled keyword. \\
    2 & \textbf{Logical Consistency} &
        Whether the lattice type, simulation procedure, and physical setup are consistent with the task specification. &
        The scenario requires \texttt{BCC} but mistakenly uses \texttt{FCC}. \\
    3 & \textbf{Parameter Rationality} &
        Whether physical parameters such as potential, temperature, and pressure are set appropriately. &
        The pressure unit in \texttt{fix npt} does not match the system’s unit style. \\
    4 & \textbf{Core Logic Accuracy} &
        Whether key computations (e.g., density, energy) are performed correctly. &
        Using volume instead of mass when calculating density. \\
    5 & \textbf{Logical Completeness} &
        Whether the script includes all essential steps such as potential setup, thermostat, and boundary conditions. &
        Thermostat not specified or \texttt{pair\_style} missing. \\
    6 & \textbf{Code Completeness} &
        Whether the script contains a complete simulation workflow (e.g., \texttt{run}, output). &
        Missing the \texttt{run} command or output directives. \\
    7 & \textbf{Result Validity} &
        Whether the script can run successfully without abnormal termination (e.g., NaN values, empty output). &
        \texttt{log} file shows ``lost atoms'' or \texttt{dump} file is empty. \\
    8 & \textbf{Physical Soundness} &
        Whether the temporal evolution of temperature, energy, or pressure follows physical laws. &
        Target temperature is 300~K, but results diverge to 3000~K. \\
    \bottomrule
  \end{tabularx}
  }
\end{table*}
\FloatBarrier

\section{Datasets Details}\label{secA3}

A comprehensive statistical analysis was performed on all datasets referenced in this work, as presented in Table~\ref{tab:datasets_statistic} and Table~\ref{tab:datasets_statistic2}
\begin{table}[ht]
  \centering

  \renewcommand{\arraystretch}{1.25} 

  \caption{Statistics of training and evaluation datasets.}
  \label{tab:datasets_statistic}
  
  \begin{tabular}{l r}
    \toprule
    \textbf{Dataset Name} & \textbf{Number / Tokens} \\
    \midrule
    \multicolumn{2}{c}{\textit{Training Datasets}} \\
    \midrule
    MD-Knowledge (\#Samples) & 17,808 \\
    MD-Knowledge (Tokens) & 10,865,191 \\
    MD-InstructQA (\#Samples) & 27,346 \\
    MD-CodeGen (\#Samples) & 4,253 \\
    \midrule
    \multicolumn{2}{c}{\textit{Evaluation Datasets}} \\
    \midrule
    MD-KnowledgeEval (\#Samples) & 336 \\
    LAMMPS-SyntaxEval (\#Samples) & 368 \\
    LAMMPS-CodeGenEval (\#Samples) & 566 \\
    \bottomrule
  \end{tabular}

  \vspace{1.5em} 

  \caption{Distribution of question types and difficulty levels in evaluation datasets.}
  \label{tab:datasets_statistic2}
    \small
  \resizebox{0.7\textwidth}{!}{ 
  \begin{tabular}{lcccc}
    \toprule
    \textbf{Dataset} & \textbf{Single} & \textbf{Multiple} & \textbf{Fill-in} & \textbf{Open-QA} \\
    \midrule
    MD-KnowledgeEval & 151 & 85 & 63 & 37 \\
    LAMMPS-SyntaxEval & 134 & 81 & 81 & 28 \\
    \midrule
    \textbf{Total} & 285 & 166 & 144 & 65 \\
    \bottomrule
  \end{tabular}
  }

  \vspace{0.5em} 

  \resizebox{0.7\textwidth}{!}{ 
  \begin{tabular}{lccc}
    \toprule
    \textbf{Dataset} & \textbf{Easy} & \textbf{Medium} & \textbf{Hard} \\
    \midrule
    MD-KnowledgeEval & 83 & 166 & 87 \\
    LAMMPS-SyntaxEval & 61 & 164 & 99 \\
    \midrule
    \textbf{Total} & 144 & 330 & 186 \\
    \bottomrule
  \end{tabular}
  }
\end{table}
\FloatBarrier

\lstdefinestyle{lammps}{
  language={},                 
  basicstyle=\ttfamily\scriptsize,
  keywordstyle=\color{blue!70!black},
  commentstyle=\color{green!50!black},
  stringstyle=\color{orange!70!black},
  numbers=none,
  frame=single,
  rulecolor=\color{black!40},
  backgroundcolor=\color{black!3},
  breaklines=true,
  breakatwhitespace=true,
  showstringspaces=false,
  tabsize=2,
  xleftmargin=6pt,
  xrightmargin=6pt,
  aboveskip=6pt,
  belowskip=6pt
}

\section{Case Study}\label{sec:case-study}
\begin{tcolorbox}[
  colback=green!5!white,
  colframe=green!80!black,
  boxrule=0.8pt,
  rounded corners=all,
  arc=3pt,
  colbacktitle=green!90!black,
  coltitle=white,
  title={\quad \textsc{Case Trajectory in MDAgent2-RUNTIME.}},
  fonttitle=\sffamily,
  fontupper=\rmfamily\scriptsize,
  colupper=black!85,
  breakable,
]
\begin{tcolorbox}[
    colback=red!70!black!70!,
    colframe=red!70!black!70!,
    boxrule=0.5pt,
    center,
    fontupper=\footnotesize,
    halign=center,
    valign=center,
    colupper=white!90!,
    height=5mm,
    left=2pt, right=2pt
]
User Task
\end{tcolorbox}
Using LAMMPS to simulate the melting process of a Cu–Ni nanoparticle.
\begin{tcolorbox}[
    colback=blue!70!black!70!,
    colframe=blue!70!black!70!,
    boxrule=0.5pt,
    center,
    fontupper=\footnotesize,
    halign=center,
    valign=center,
    colupper=white!90!,
    height=5mm,
    left=2pt, right=2pt
]
Step 1 - Code Generator
\end{tcolorbox}
\noindent\textbf{[Write Code]} 
\begin{lstlisting}[style=lammps]
lammps_code'''
# ---[Omitted] ---
pair_style eam/alloy
pair_coeff * * CuNi.eam Cu Ni
# ---[Omitted] ---
'''
\end{lstlisting}
\noindent\textbf{[Tool call -- Potential Tool]}  
The potentials checking tool scans the script and detects that the referenced file
\texttt{CuNi.eam} is not available in the local environment. A compatible alloy potential file \texttt{CuNi.eam.alloy} is recommended based on
material composition and filename similarity.

\noindent\textbf{[Next Action -- Generate]}  

\begin{tcolorbox}[
    colback=blue!70!black!70!,
    colframe=blue!70!black!70!,
    boxrule=0.5pt,
    center,
    fontupper=\footnotesize,
    halign=center,
    valign=center,
    colupper=white!90!,
    height=5mm,
    left=2pt, right=2pt
]
Step 2 - Code Generator
\end{tcolorbox}
\noindent\textbf{[Write Code]} 
\begin{lstlisting}[style=lammps]
lammps_code'''
# ---[Omitted] ---
pair_style eam/alloy
pair_coeff * * CuNi.eam.alloy Cu Ni
# ---[Omitted] ---
'''
\end{lstlisting}
\noindent\textbf{[Tool call -- Potential Tool]}
The potentials checking tool verifies that the specified file
\texttt{CuNi.eam.alloy} is available and compatible with the Cu--Ni alloy system.

\noindent\textbf{[Tool call -- Syntax Tool]}
The syntax checking tool confirms that the revised script conforms to valid
LAMMPS syntax rules and is free of structural errors.

\noindent\textbf{[Next Action -- Run]}

\begin{tcolorbox}[
    colback=orange!70!black!70!,
    colframe=orange!70!black!70!,
    boxrule=0.5pt,
    center,
    fontupper=\footnotesize,
    halign=center,
    valign=center,
    colupper=white!90!,
    height=5mm,
    left=2pt, right=2pt
]
Step 3 - Code Runner
\end{tcolorbox}
\noindent\textbf{[Run Lammps]} 
\begin{lstlisting}[style=lammps]
{
  status: "Execution completed successfully",
  error: "None",
  runtime: "<elapsed_time>",
  summary_info_by_rule: {
    ...[Omitted]...
  }
}
'''
\end{lstlisting}
\noindent\textbf{[Generated Artifacts]}
The following result files are produced during execution:
\begin{itemize}
\item \texttt{log.lammps}
\end{itemize}

\noindent\textbf{[Next Action -- Eval]}

\begin{tcolorbox}[
    colback=teal!40!black!65!,
    colframe=teal!40!black!65!,
    boxrule=0.5pt,
    center,
    fontupper=\footnotesize,
    halign=center,
    valign=center,
    colupper=white!90!,
    height=5mm,
    left=2pt, right=2pt
]
Step 4 - Result Evaluator
\end{tcolorbox}
\noindent\textbf{[Eval]} 
\begin{lstlisting}[style=lammps]
{
"final_score": 4,
"module_detail":[...], #[Omitted]
"penalty_detail":[...] #[Omitted]
}
\end{lstlisting}
\noindent\textbf{[Score Assessment]}  
The obtained score falls below the predefined acceptance threshold.  
The result is therefore marked as \emph{not acceptable}, and further code revision
is required.

\noindent\textbf{[Next Action -- Generate]}

\begin{tcolorbox}[
    colback=blue!70!black!70!,
    colframe=blue!70!black!70!,
    boxrule=0.5pt,
    center,
    fontupper=\footnotesize,
    halign=center,
    valign=center,
    colupper=white!90!,
    height=5mm,
    left=2pt, right=2pt
]
Step 5 - Code Generator
\end{tcolorbox}
\noindent\textbf{[Write Code]} 
\begin{lstlisting}[style=lammps]
lammps_code'''
# ---[Omitted] ---
pair_style eam/alloy
pair_coeff * * CuNi.eam.alloy Cu Ni
# ---[Omitted] ---
'''
\end{lstlisting}
\noindent\textbf{[Tool call -- Potential Tool]}
The potentials checking tool verifies that the specified file
\texttt{CuNi.eam.alloy} is available and compatible with the Cu--Ni alloy system.

\noindent\textbf{[Tool call -- Syntax Tool]}
The syntax checking tool confirms that the revised script conforms to valid
LAMMPS syntax rules and is free of structural errors.

\noindent\textbf{[Next Action -- Run]}
\begin{tcolorbox}[
    colback=orange!70!black!70!,
    colframe=orange!70!black!70!,
    boxrule=0.5pt,
    center,
    fontupper=\footnotesize,
    halign=center,
    valign=center,
    colupper=white!90!,
    height=5mm,
    left=2pt, right=2pt
]
Step 6 - Code Runner
\end{tcolorbox}

\begin{tcolorbox}[
    colback=teal!40!black!65!,
    colframe=teal!40!black!65!,
    boxrule=0.5pt,
    center,
    fontupper=\footnotesize,
    halign=center,
    valign=center,
    colupper=white!90!,
    height=5mm,
    left=2pt, right=2pt
]
Step 7 - Result Evaluator
\end{tcolorbox}
\begin{tcolorbox}[
    colback=blue!70!black!70!,
    colframe=blue!70!black!70!,
    boxrule=0.5pt,
    center,
    fontupper=\footnotesize,
    halign=center,
    valign=center,
    colupper=white!90!,
    height=5mm,
    left=2pt, right=2pt
]
Step 8 - Code Generator
\end{tcolorbox}
\begin{tcolorbox}[
    colback=orange!70!black!70!,
    colframe=orange!70!black!70!,
    boxrule=0.5pt,
    center,
    fontupper=\footnotesize,
    halign=center,
    valign=center,
    colupper=white!90!,
    height=5mm,
    left=2pt, right=2pt
]
Step 9 - Code Runner
\end{tcolorbox}

\begin{tcolorbox}[
    colback=teal!40!black!65!,
    colframe=teal!40!black!65!,
    boxrule=0.5pt,
    center,
    fontupper=\footnotesize,
    halign=center,
    valign=center,
    colupper=white!90!,
    height=5mm,
    left=2pt, right=2pt
]
Step 10 - Result Evaluator
\end{tcolorbox}
\begin{tcolorbox}[
    colback=red!70!black!70!,
    colframe=red!70!black!70!,
    boxrule=0.5pt,
    center,
    fontupper=\footnotesize,
    halign=center,
    valign=center,
    colupper=white!90!,
    height=5mm,
    left=2pt, right=2pt
]
Step 10 - END
\end{tcolorbox}
\noindent\textbf{[Final Answer]} 
\begin{lstlisting}[style=lammps]
lammps_code'''
units           metal
boundary        p p p
atom_style      atomic

variable        A0 equal 3.589
lattice         fcc ${A0}

region          mybox block 0 40 0 40 0 40
create_box      2 mybox                     # 2 atom types

# Nanoparticle (sphere)
region          CuNi_nano sphere 20 20 20 2
create_atoms    1 region CuNi_nano
set             type 1 type/fraction 2 0.5 7777

# Atomic masses
mass            1 63.54600000               # Cu
mass            2 58.69340000               # Ni

pair_style      eam/alloy
pair_coeff      * * CuNi.eam.alloy Cu Ni

# Output initial structure
write_data      nanoparticle.cif
run             0

thermo          1000

variable        j loop 0 20
label           loop_j

# Temperature definition
variable        temperature equal 900 + 10*${j}
variable        T           equal temp
variable        Eatom       equal etotal/atoms

fix             1 all nvt temp ${temperature} ${temperature} 0.1
run             50000
unfix           1

fix             2 all ave/time 100 5 1000 v_T v_Eatom \
                file data_ave${temperature}.txt
dump            1 all atom 5000 fusion_${temperature}.atom

fix             1 all nvt temp ${temperature} ${temperature} 0.1
run             500000

unfix           1
undump          1
unfix           2

next            j
jump            SELF loop_j
'''
\end{lstlisting}

\end{tcolorbox}

\end{appendices}

\end{document}